\documentclass[sigconf]{acmart}

\usepackage{amsfonts}
\usepackage{multirow}
\usepackage{subcaption}
\usepackage{makecell}
\usepackage{balance}
\hypersetup{
colorlinks=true,
linkcolor=blue,
urlcolor=cyan
}
\AtBeginDocument{%
  \providecommand\BibTeX{{%
    \normalfont B\kern-0.5em{\scshape i\kern-0.25em b}\kern-0.8em\TeX}}}

\setcopyright{acmcopyright}
\copyrightyear{2023}
\acmYear{2023}

\setcopyright{acmlicensed}\acmConference[MM '23]{Proceedings of the 31st ACM International Conference on Multimedia}{October 29-November 3, 2023}{Ottawa, ON, Canada}
\acmBooktitle{Proceedings of the 31st ACM International Conference on Multimedia (MM '23), October 29-November 3, 2023, Ottawa, ON, Canada}
\acmPrice{15.00}
\acmDOI{10.1145/3581783.3612269}
\acmISBN{979-8-4007-0108-5/23/10}

\begin{document}


\title{Exploiting Time-Frequency Conformers for Music Audio Enhancement}


\author{Yunkee Chae}
\email{yunkimo95@snu.ac.kr}
\orcid{0009-0000-8073-3754}
\affiliation{
    \institution{Seoul National University}
    \city{Seoul}
    \country{Republic of Korea}
}

\author{Junghyun Koo}
\email{dg22302@snu.ac.kr}
\affiliation{
    \institution{Seoul National University}
    \city{Seoul}
    \country{Republic of Korea}
}

\author{Sungho Lee}
\email{sh-lee@snu.ac.kr}
\affiliation{
    \institution{Seoul National University}
    \city{Seoul}
    \country{Republic of Korea}
}

\author{Kyogu Lee}
\email{kglee@snu.ac.kr}
\affiliation{
    \institution{Seoul National University}
    \city{Seoul}
    \country{Republic of Korea}
}

\renewcommand{\shortauthors}{Yunkee Chae, Junghyun Koo, Sungho Lee, \& Kyogu Lee}

\begin{abstract}
With the proliferation of video platforms on the internet, recording musical performances by mobile devices has become commonplace. 
However, these recordings often suffer from degradation such as noise and reverberation, which negatively impact the listening experience.
Consequently, the necessity for music audio enhancement (referred to as music enhancement from this point onward), involving the transformation of degraded audio recordings into pristine high-quality music, has surged to augment the auditory experience.
To address this issue, we propose a music enhancement system based on the Conformer architecture that has demonstrated outstanding performance in speech enhancement tasks.
Our approach explores the attention mechanisms of the Conformer and examines their performance to discover the best approach for the music enhancement task.
Our experimental results show that our proposed model achieves state-of-the-art performance on single-stem music enhancement.
Furthermore, our system can perform general music enhancement with multi-track mixtures, which has not been examined in previous work. 
Audio samples enhanced with our system are available at: 
\url{https://tinyurl.com/smpls9999}
\end{abstract}






\maketitle

\section{Introduction}
The rapid expansion of social media platforms on the internet has led to unprecedented accessibility to a large number of videos and audio recordings captured by unprofessional devices, such as smartphones. 
Despite the convenience of these recordings, their quality often suffers due to various factors, such as background noise, reverberation, and frequency responses of the microphone devices.
In particular, live performance recordings on platforms like YouTube typically lack the audio quality expected from professionally recorded studio tracks.
As a result, the auditory experience of these recordings is significantly diminished, creating a growing demand for music enhancement techniques that can restore the original quality of the audio.
Therefore, post-processing tools that can enhance distorted musical recordings to resemble studio-quality audio are essential for an improved personal listening experience. 

\begin{figure}[t]
    \centering
    \includegraphics[width=1.0\columnwidth]{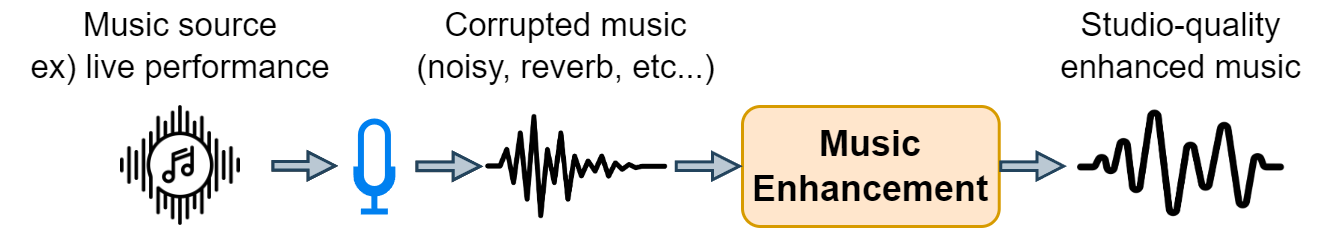}
    \caption{Framework overview of music enhancement task.}
    \label{fig:task_overview}
\end{figure}



Music enhancement, which involves transforming distorted audio recordings into pristine high-quality music, plays a crucial role in augmenting the auditory experience for listeners.
Recent studies have sought to address the challenge of enhancing audio quality in such recordings, with many leveraging deep learning techniques to achieve impressive results.
Kandapal \textit{et al.} \cite{refme} made significant contributions to the field by enhancing music recordings using the Pix2Pix \cite{pix2pix} and Diffwave \cite{diffwave} approaches in their study, a pioneering work in music enhancement.
While their work has been influential, it focused primarily on single-stem music datasets, which led to certain limitations.
For instance, their models were not designed to handle general multi-track music, as they  were trained on pre-defined classes of instruments using the Medley-solos-DB \cite{medleysolosdb} dataset, which limited their ability to generate waveforms of other general instruments.

Schaffer \textit{et al.} \cite{mss_enhance} investigated music enhancement to improve the performance of music source separation models using generative modeling. Their proposed post-processor effectively enhanced the estimates of bass and drums, as demonstrated through subjective evaluations.
However, their method was primarily focused on enhancing the outputs of the source separation model, which are expected to be a single stem. 

Despite the advancements made by these works, there remains room for further exploration in multi-track music enhancement.
In particular, the development of methods that can effectively handle a diverse range of instruments and musical genres would be a valuable addition to the field.
Moreover, creating systems capable of enhancing multi-track mixtures could unlock new possibilities for music enhancement tasks, enabling users to experience high-quality audio across a broader spectrum of musical compositions. 

Accordingly, we propose a new system for music enhancement tasks based on the widely-used Conformer architecture \cite{Conformer}. 
We introduce new modules called TF-Conformers, which adopt attention mechanisms for time-frequency representations of musical signals.
Our contributions are as follows:
\begin{itemize}
    \item The proposed system achieves state-of-the-art performance in single-stem music enhancement tasks.
    \item We extend the validation of our system by investigating its applicability to multi-track mixture music enhancement, an area that has not yet been extensively explored in the literature.
    \item We explore new TF-Conformer modules, incorporating attention mechanisms, and evaluate their effectiveness through several ablation studies.
\end{itemize}

The remainder of this paper is organized as follows: Section \ref{sec:related_works} provides an overview of recent work in the research field related to music enhancement, focusing on deep learning-based approaches.
Section \ref{sec:methods} details our proposed Conformer-based music enhancement system, including the architecture and attention mechanisms employed.
Section \ref{sec:experiments} presents the dataset used in our experiments, implementation details, and evaluation metrics. 
Section \ref{sec:results} reports the results of both objective and subjective evaluation from our experiments.
Finally, Section \ref{sec:conclusions} offers concluding remarks, summarizes the contributions of this paper, and discusses future work in this area.

\begin{figure*}[t]
    \centering
    \includegraphics[width=\textwidth]{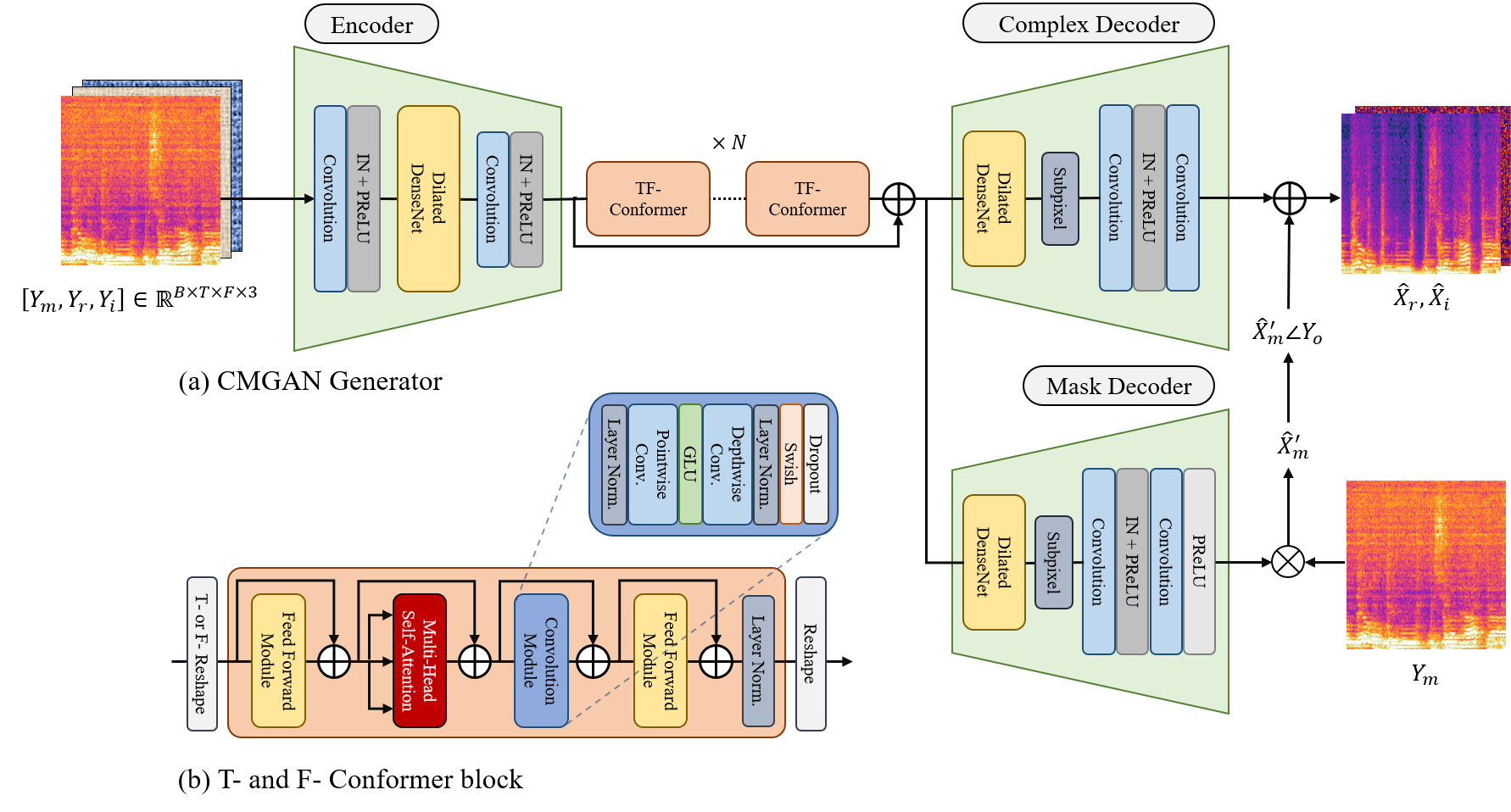}
    \caption{Overall enhancement process of the proposed system based on CMGAN generator \cite{cmganlong}.
    The model takes a compressed spectrogram of noisy music, $Y\in\mathbb{C}^{B\times T\times F}$, as input and outputs the real and imaginary parts of the compressed spectrogram $\hat{X}$ for the enhanced music, denoted by $\hat{X}_r$ and $\hat{X}_i$.
    These outputs are then compared to the ground truth clean compressed spectrogram $X$ using various loss functions. 
    $Y_m$, $Y_r$, and $Y_i$ represent the magnitude spectrogram, and the real and imaginary parts of $Y$, respectively.
    The mask decoder estimates a mask for the noisy magnitude spectrogram, while the complex decoder estimates the residuals of the real and imaginary parts to derive the enhanced spectrogram.
    }
    \label{fig:overall_cmgan}
    \vspace{-1mm}
\end{figure*}

\section{Related Works}\label{sec:related_works}
Kandpal \textit{et al.} \cite{refme} proposed the \textit{Mel2Mel + Diffwave} framework for music enhancement tasks, wherein Mel2Mel and Diffwave models are based on \cite{pix2pix} and \cite{diffwave}, respectively.
Mel2Mel is employed to enhance the mel-spectrogram of distorted music while Diffwave serves as a vocoder, converting the mel-spectrogram to waveform.
They trained \textit{Mel2Mel + Diffwave} models independently or jointly, with each approach offering distinct benefits.
Independent training promotes robustness to artifacts in the enhanced mel-spectrogram, as the Diffwave vocoder is trained exclusively on clean mel-spectrograms.
On the other hand, joint training yields a higher FAD score \cite{fad}, which the authors argue is closely related to human perception.

In the domain of music source separation, Schaffer \textit{et al.} \cite{mss_enhance} presented the Make it Sound Good (MSG) post-processor to enhance the output of the music source separation system. They are tailored to counter the perceptual deficiencies of music source separation systems, such as the emergence of superfluous noise or the elimination of harmonics.
The authors conducted their work utilizing generative modeling, with the generator derived from an architecture akin to Demucs v2 \cite{demucsv2} and the discriminator adopted from HiFi-GAN \cite{hifigan}, optimized with the losses of LSGAN \cite{lsgan} and deep feature matching loss \cite{larsen2016autoencoding}.
The authors employed their post-processing model on state-of-the-art music source separators, including a separator not seen during training.
Their findings demonstrated that MSG can significantly enhance the quality of the source estimates.

The field of audio enhancement has predominantly focused on speech enhancement \cite{segan, dcunet, dccrn, trunet}.
Especially, recent deep learning-based speech enhancement models have adopted self-attention mechanisms \cite{allyouneed} due to their ability to capture long-range dependencies in sequential features.
In particular, various methods apply self-attention to both the frequency-axis and time-axis in different configurations, which we refer to as \textit{TF-self-attentions}. 
These methods have demonstrated the effectiveness of such attention mechanisms in enhancing audio quality \cite{sa_cascade, sa_parallel_cycle, sa_parallel_joint, mtfaa}.
Meanwhile, Conformer-based models have achieved state-of-the-art performances in speech enhancement tasks \cite{seConformer, dfConformer, uformer, cmganlong}.
Specifically, CMGAN \cite{cmganlong} and Uformer \cite{uformer} utilize time- and frequency-Conformers, which are applied to the time and frequency axes of the spectrogram.
To the best of our knowledge, no existing works incorporate other TF-self-attention methods into Conformer-based models. 
As such, we propose a novel approach that integrates various TF-self-attention mechanisms into Conformer-based models, which we term the \textit{TF-Conformer module}.
This approach aims to enhance music quality, ultimately achieving state-of-the-art performance.

\section{Methods}\label{sec:methods}
In this study, we concentrate on the general music enhancement problem, aiming to improve the quality of music recordings, especially those that include not merely single-stem recordings but also multi-track mixtures. 
To achieve this, we investigate the use of TF-Conformer modules, leveraging the encoder and decoder components of the CMGAN generator \cite{cmganlong}, as illustrated in Fig. \ref{fig:overall_cmgan}.
In this section, we provide a brief overview of the CMGAN generator architecture and introduce our various TF-Conformer modules.
Lastly, we discuss the loss functions utilized to train our models.

\subsection{CMGAN Generator}

\begin{figure*}[t]
    \centering
    \includegraphics[width=\textwidth]{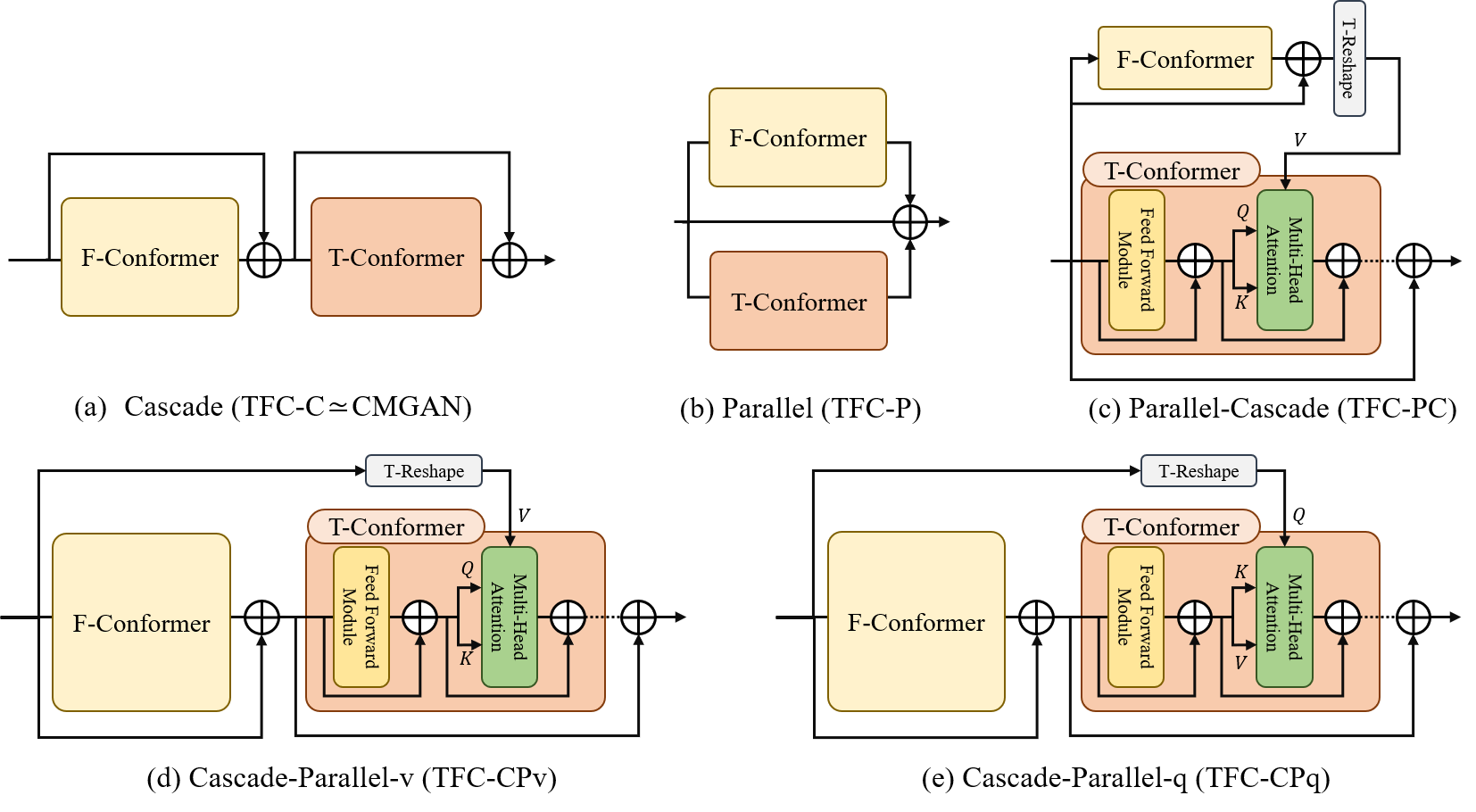}
    \setlength{\abovecaptionskip}{-10pt}
    \caption{Description of the proposed TF-Conformer modules. We evaluate the enhancement performance of each module. $Q$,$K$, and $V$ denote the query, key, and value of the multi-head attention, respectively. T- and F-Conformers correspond to the structures depicted in Fig. \ref{fig:overall_cmgan}-(b), with the distinction that in (c), (d), and (e), the multi-head self-attention is substituted with multi-head attention.}
    \label{fig:tfConformers}
    \vspace{-1mm}
\end{figure*}

Our method focuses on enhancing music within the time-frequency (TF) domain, particularly by employing the compressed magnitude spectrogram representation and incorporating phase information as well.
This allows us to handle complex harmonic structures and polyphonic sound sources effectively. 
The CMGAN generator takes compressed magnitude spectrogram $Y_m(t, f)$, and real $Y_r(t, f)$ and imaginary parts $Y_i(t,f)$ of the compressed complex spectrogram $Y$ as the input, which is formulated as follows:
\begin{equation}\label{eq:noisyeq}
    Y=|Y_o|^ce^{j\angle{Y_o}}=Y_me^{j\angle{Y_o}}=Y_r+jY_i
\end{equation}
where $Y_o\in \mathbb{C}^{T\times F}$ denotes the complex spectrogram of low-quality music inputs converted by the short-time Fourier transform (STFT). Here, $T$ and $F$ denote the number of time frames and frequency bins, respectively.

The CMGAN encodes the spectrogram with a dilated DenseNet \cite{densenet} in between two convolutional layers, where the latter layer halves the frequency dimension. 
Each convolutional layer is followed by instance normalization \cite{instancenorm} and PReLU \cite{prelu} subsequently.
The dilated DenseNet has dilation factors of 1, 2, 4, and 8 for each convolutional layer. 
Based on our observations, we found that incorporating a residual connection between the encoder's output and the decoder's input results in faster convergence.
Therefore, we added this residual connection to the original CMGAN generator's encoder to enhance its performance.

The CMGAN generator consists of two decoders: the magnitude mask decoder and the complex residual decoder.
The mask decoder calculates the mask for element-wise multiplication with the input low-quality magnitude spectrogram.
Meanwhile, the complex decoder computes real and imaginary residuals to be added to the masked magnitude spectrogram along with the original input phase, producing the final complex spectrogram output.
Both decoders employ dilated DenseNets and a subpixel convolution layer \cite{subpixel} to upsample the frequency dimension.
Subsequent convolutional layers adjust the number of output channels to 1 and 2, respectively, followed by instance normalization and an additional convolutional layer.
The PreLU activation function \cite{prelu} is applied exclusively to the mask decoder.

The final output $(\hat{X}_r, \hat{X}_i)$ of the CMGAN generator is calculated as follows:
\begin{equation}
    \hat{X}_r=\hat{X}'_m\cos{(\angle{Y_o})}+\hat{X}'_r, \quad \hat{X}_i=\hat{X}'_m\sin{(\angle{Y_o})}+\hat{X}'_i
\end{equation}
where $\hat{X}'_m$ is the input magnitude spectrogram masked by the output of the mask decoder and $\hat{X}'_r$, $\hat{X}'_i$ denotes the outputs of the complex decoder.

\subsection{TF-Conformer}


TF-self-attention mechanisms have been employed by numerous speech enhancement tasks \cite{sa_cascade, sa_parallel_cycle, sa_parallel_joint, mtfaa}.
We explore various attention mechanisms in the context of Conformers to examine which method is optimal for music enhancement.


We introduce our proposed \textit{TF-Conformers} (TFC) in Fig. \ref{fig:tfConformers}, which are based on the conformers depicted in \ref{fig:overall_cmgan}-(b).
\textit{T-Conformer} and \textit{F-Conformer} transform the input dimensions to $\mathbb{R}^{(BF)\times T\times C}$ and $\mathbb{R}^{(BT)\times F \times C}$, respectively, and the final reshape block restores the output to its original shape, matching the input dimension.
These Conformers are designed to model the sequential features of the time and frequency in spectrograms, respectively.
The T-Conformer focuses on capturing the sequential features along the time axis, treating the frequency axis as a global feature for each time frame.
This enables the T-Conformer to effectively model temporal dependencies and dynamics present in the spectrogram.
Similarly, the F-Conformer is tailored to model sequential features along the frequency axis, considering the time axis as a global feature for each frequency frame.
This approach allows the F-Conformer to capture spectral patterns and dependencies within the spectrogram.
By separately modeling time and frequency features, the T- and F-Conformers can better comprehend and represent the complex structure of spectrograms, ultimately leading to enhanced performance in music enhancement tasks.

\begin{table*}[t!]
    \centering{}
    \setlength\tabcolsep{3.2pt}
    \begin{tabular}{l ccccc ccccc ccccc}
    \toprule
    \multirow{2}{*}{Model} & \multicolumn{5}{c}{Medley-solos-DB (\textit{solo})} & \multicolumn{5}{c}{Medley-solos-DB (piano-only)} & \multicolumn{5}{c}{MUSDB18}\\ \cmidrule(lr){2-6} \cmidrule(lr){7-11} \cmidrule(lr){12-16}
    & fwSNR $\uparrow$ & MRS $\downarrow$ & L1 $\downarrow$ & SDR $\uparrow$ & FAD $\downarrow$
    & fwSNR $\uparrow$ & MRS $\downarrow$ & L1 $\downarrow$ & SDR $\uparrow$ & FAD $\downarrow$ 
    & fwSNR $\uparrow$ & MRS $\downarrow$ & L1 $\downarrow$ & SDR $\uparrow$ & FAD $\downarrow$ \\ \midrule
    LQ  & 5.68 & 1.75 & 2.69 & -2.69 & 5.70
        & 6.88& 1.98& 3.09& -2.90 & 6.87
        & 6.72 & 1.64 & 2.36 & -3.39 & 11.30\\ 
    \textit{M+D} \cite{refme}  & 8.03& 1.45& 2.24 & -2.31&3.68 
        & 8.96& 1.44& 2.56& -1.33& 4.05
        & - & - & - & - & -      \\ \midrule
    TFC-C   & 10.48 & 1.05 & 1.74 & 3.91 & 0.77 
        & 11.30 & 1.06 & 1.65 & 3.93 & 0.86
        &12.49 & 0.83 & 1.28 & 4.48 & 0.65\\
    TFC-P   & 10.49 & 1.04 & 1.72 & 3.77 & 0.65
        & 11.03 & 1.03 & 1.61 & 4.09 & 0.47
        & \textbf{12.73} & \textbf{0.81} & 1.26 & 4.52 & 0.77\\
    TFC-PC  & 10.10 & 1.05 & 1.71 & 3.69 & 1.02 
        & \textbf{11.37} & \textbf{0.99} & \textbf{1.57} & 4.27 & 0.50
        & 12.22 & 0.83 & 1.28 & 4.43 & 0.77\\
    TFC-CPv & \textbf{10.69} & 1.01 & 1.65 & 3.91 & 0.68
        & 11.21 & 1.03 & 1.65 & 4.02 & \textbf{0.30}
        & 12.57 & 0.84 & 1.30 & 4.21 & 0.66\\
    TFC-CPq & 10.51& \textbf{0.99} & \textbf{1.63} & \textbf{4.13} & \textbf{0.62}
        & 11.07 & 1.03 & 1.62 & \textbf{4.43} & 0.45
        & 12.68  &\textbf{0.81} & \textbf{1.25} & \textbf{4.66} & \textbf{0.53} \\  \midrule
    TFC-CPq (M) & 9.73 & 1.16 & 1.93 & 3.90 & 2.38
        &10.99 & 1.36 & 2.28 & 4.10 & 3.26
        & - & - & - & - & - \\ 
        \bottomrule
    \end{tabular}
    \caption{Objective evaluation results for \textit{solo} and MUSDB18 dataset, where LQ and \textit{M+D} denotes low-quality musical recordings and \textit{Mel2Mel + Diffwave}, respectively. 
    On the Medley-solos-DB (with only piano) columns, we reported the evaluations trained only on piano samples of the \textit{solo} dataset.
    In the TFC-CPq (M) row, we report the evaluations of enhanced samples of the \textit{solo} dataset from proposed models trained on the MUSDB18 dataset.
    }
    \vspace{-1mm}
    \label{table:obj solo musdb}
\end{table*}

It should be noted that the Cascade TF-Conformer (TFC-C) is similar to the two-stage Conformer block \cite{cmganlong} except that it applies the F-Conformer before the T-Conformer.
Parallel TF-Conformer (TFC-P) processes input through two parallel branches, similar to Adaptive Time-Frequency Attention (ATFA) module in \cite{sa_parallel_cycle}.
Parallel-Cascade TF-Conformer (TFC-PC) takes inspiration from the Axial Self-Attention (ASA) module of \cite{mtfaa}. We have applied the methodology of ASA to the multi-head attention (MHA) module within the Conformer module.
In this module, the MHA module of the T-Conformer receives the value input from the output of the F-Conformer, as depicted in Fig. \ref{fig:tfConformers}-(c).
Furthermore, we propose the Cascade-Parallel module, including Cascade-Parallel-value (TFC-CPv) module and the Cascade-Parallel-query (TFC-CPq) module.
The TFC-CPv processes input sequentially through the F- and T-Conformers, where the MHA of the T-Conformers receives the value directly from the input of the TFC-CPv module.
The TFC-CPq architecture is designed akin to the TFC-CPv; however, it acquires the query input for the MHA directly from the input of the TFC module, while key and value inputs are obtained from the output of the F-Conformer.   
This configuration can align more closely with the conceptual underpinnings of key, value, and query in the Transformer \cite{allyouneed} model, in comparison to the TFC-CPv and TFC-PC cases.
To the best of our knowledge, these methods have not been proposed in the context of TF-self-attention-based models.
It is worth noting that all of the TF-Conformers have the same number of parameters.

\subsection{Loss Functions}
We define magnitude, real and imaginary parts $X_m(t, f)$, $X_r(t,f)$, $X_i(t,f)$ of compressed spectrogram $X$ using the clean music source spectrogram $X_o\in\mathbb{C}^{T\times F}$, same as in equation (\ref{eq:noisyeq}):
\begin{equation}\label{eq:cleaneq}
    X=|X_o|^ce^{j\angle{X_o}}=X_me^{j\angle{X_o}}=X_r+jX_i \ .
\end{equation}

Meanwhile, we compute the estimates of the clean spectrogram and waveform as follows:
\begin{align}
    & \hat{X}_m=\sqrt{\hat{X}_r^2+\hat{X}_i^2},\\
    & \hat{x} = \text{iSTFT}(\hat{X}_m^{(1-c)/c}(\hat{X}_r+j\hat{X}_i))
    \label{eq:est}
\end{align}
where $\text{iSTFT}(\cdot)$ denotes the inverse short-time Fourier transform.
Subsequently, we employ a linear combination of magnitude loss, complex loss, and time loss, following the approach in \cite{cmganlong}:
\begin{align}
    & \mathcal{L}_{\text{Mag}}=\mathbb{E}_{X_m,\hat{X}_m}[\lVert X_m-\hat{X}_m \rVert^2] \nonumber\\
    & \mathcal{L}_{\text{RI}}=\mathbb{E}_{X_r,\hat{X}_r}[\lVert X_r-\hat{X}_r\rVert^2]+
        \mathbb{E}_{X_i, \hat{X}_i}[\lVert X_i-\hat{X}_i\rVert^2] \\
    & \mathcal{L}_{\text{Time}}=\mathbb{E}_{x, \hat{x}}[\lVert x-\hat{x} \rVert_1] \nonumber
\end{align}
where $x$, $\hat{x}$ denotes the clean music waveform and the estimated waveform, respectively.
Finally, the total loss is $\mathcal{L}$ is defined by:
\begin{align}
    & \mathcal{L}=\gamma_1\mathcal{L}_{\text{Mag}}+\gamma_2\mathcal{L}_{\text{RI}}+\gamma_3\mathcal{L}_{\text{Time}} \ .
\end{align}

\section{Experiments}\label{sec:experiments}

\subsection{Dataset}\label{subsec:dataset}
Our models are trained on two distinct datasets: the Medley-solos-DB dataset \cite{medleysolosdb} (\textit{solo}) for comparison with previous models, and MUSDB18 \cite{musdb18} for general multi-track music enhancement.
Initially, we use \textit{solo} dataset excluding the distorted electric guitar class, which served as the training data for \textit{Mel2Mel + Diffwave} \cite{refme}, the prior state-of-the-art model.
The dataset comprises 19,171 single-stem recordings of approximately 3 seconds each and is divided into training, validation, and test subsets containing 5,437, 2,999, and 11,281 samples, respectively.
Additionally, we compare the performance of models using a piano-only \textit{solo} dataset, which includes a total of 5,672 piano samples--2,041 for the training set, 1,022 for validation, and 2,609 for the test set.
Furthermore, we employ the MUSDB18 \cite{musdb18} dataset, consisting of 150 multi-track songs for general multi-track music enhancement.
Excluding the 50 songs designated for the test set, we randomly select 90 songs for the training set and 10 songs for the validation set.
Furthermore, we partitioned the MUSDB18 data into 3-second segments, resulting in 6800, 847, and 4166 samples for the training, validation, and test sets, respectively.
We use only the mixture files from each song and segment them to match the length of samples in the \textit{solo} dataset.

To fairly compare our approach with the \textit{Mel2Mel + Diffwave} model, we adopt the same data simulation schemes as described in \cite{refme}. 
In order to create aligned pairs of clean high-quality and corrupted low-quality music, we convolve the clean music source with room impulse responses from the DNS Challenge dataset \cite{dnschallenge}, simulating a reverberant environment.
Additionally, we introduce realistic noise from the ACE challenge dataset \cite{acechallenge}, scaled according to randomly sampled signal-to-noise ratios (SNR) between 5dB and 30dB.
We simulate various frequency responses of low-quality microphones by applying random gain between [-15dB, 15dB] to four different frequency bands ([0, 200], [200, 1000], [1000, 4000], and [4000, 8000] Hz).
In accordance with \cite{refme}, we also implement a low-cut filter to remove nearly imperceptible frequencies below 35 Hz. Additionally, we normalize the waveforms by dividing each segment by its maximum absolute value and scaling them by a factor of 0.95.

\begin{figure*}[t]
    \centering
    \includegraphics[width=\textwidth]{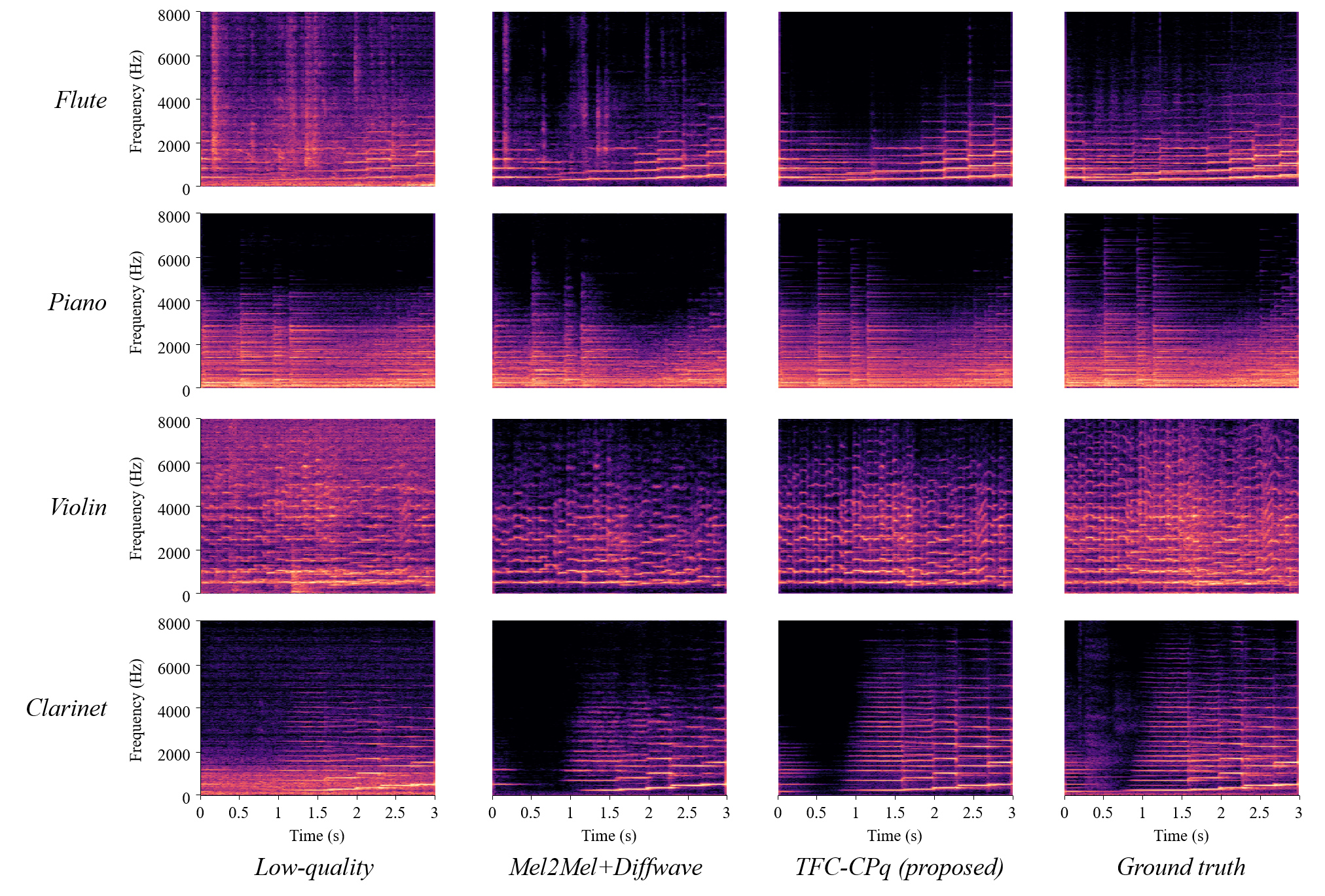}
    \setlength{\abovecaptionskip}{-10pt}
    \caption{Comparison between the \textit{Mel2Mel+Diffwave} \cite{refme} and the proposed system with four different individual instruments in the Medley-solos-DB dataset.    
    \textit{Ground truth} refers to the clean audio within the dataset, while \textit{Low-quality} denotes the synthesized noisy audio containing noise and reverberation.
    Our system more effectively denoises background noise, recovers high-frequency harmonics, and exhibits a more precise reconstruction of spectral features compared to the \textit{Mel2Mel+Diffwave}.
    }
    \label{fig:spec_solo}
    \vspace{-1mm}
\end{figure*}

\subsection{Implementation Details}
We perform STFT with a Hamming window using an FFT size of 1024 samples and 75\% overlap.
All waveforms are processed at a 16 kHz sampling rate.
For \textit{Mel2Mel + Diffwave}, we follow the configuration described in \cite{refme}.
Although \cite{refme} proposes several training schemes for \textit{Mel2Mel + Diffwave}, we employ the independent training method, which trains Mel2Mel and Diffwave separately.
We use pre-trained weights from the official repository\footnote{\url{https://github.com/nkandpa2/music_enhancement}} to estimate enhancement performance on samples from the \textit{solo} dataset.
For the CMGAN generator bottleneck, we apply two TF-Conformer modules to each of our proposed models.
The compression exponent, denoted as $c$, is set to 0.3 for magnitude spectrograms. Meanwhile, the values for $\gamma_1$, $\gamma_2$, and $\gamma_3$ in the loss function are set to 0.15, 0.85, and 0.1, respectively, 
based on a grid search.
All the proposed models are trained for 50 epochs on both the \textit{solo} and MUSDB18 datasets.
For the piano-only \textit{solo} dataset, the training epoch is set to 250.
The optimization process employs the AdamW optimizer with a learning rate of 0.00005 and a batch size of 8.

\begin{figure*}[t]
    \centering
    \includegraphics[width=\textwidth]{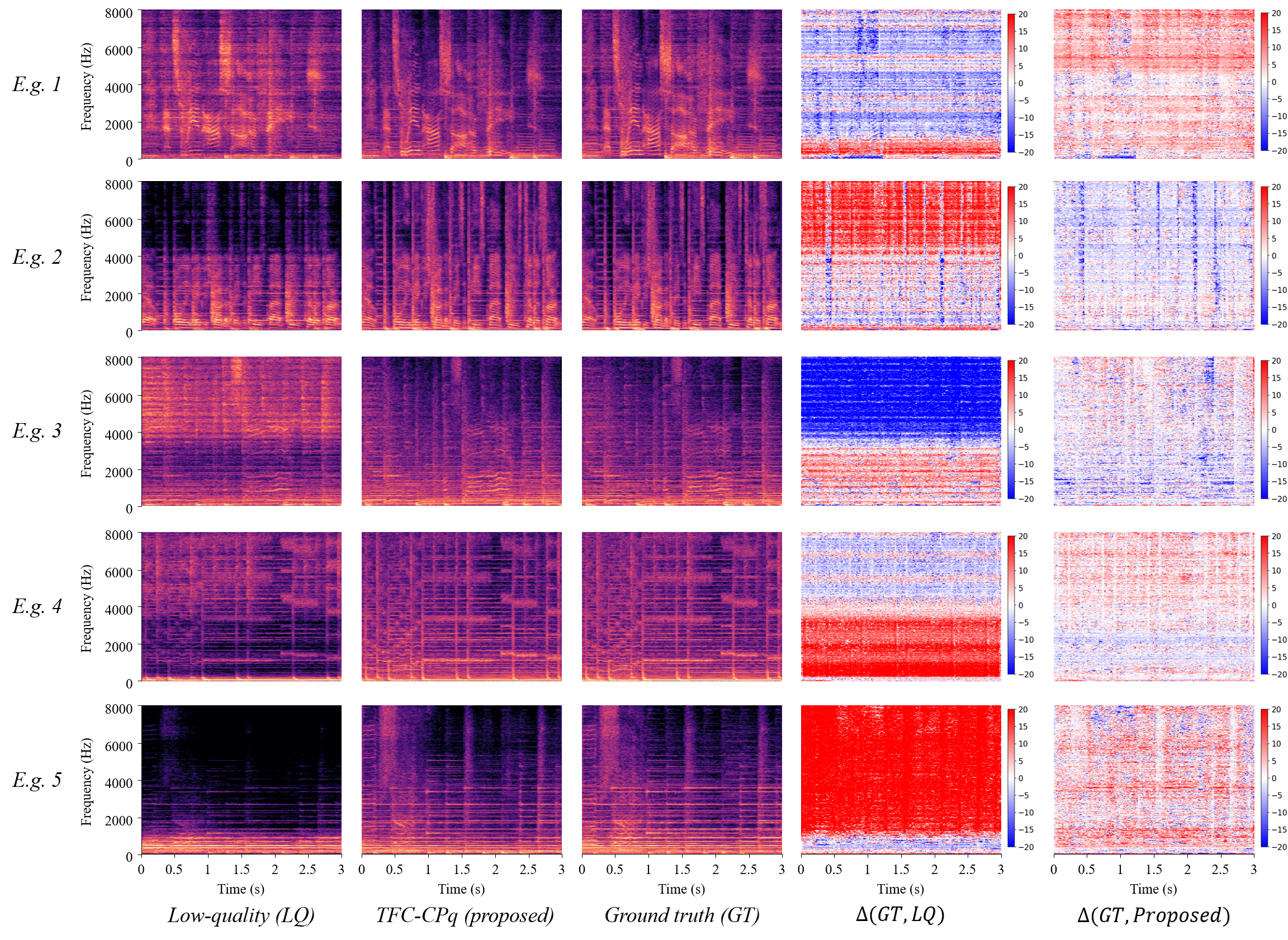}
    \setlength{\abovecaptionskip}{-10pt}
    \caption{Magnitude spectrogram results of MUSDB18 dataset enhanced by the proposed system.
    $\Delta(GT, LQ)$ refers to the difference between the ground truth and low-quality music spectrograms, while $\Delta(GT, Proposed)$ denotes the difference between the ground truth and the enhanced music spectrogram.
    In other words, within the $\Delta(GT, \cdot)$ representation, red bins signify the absence of expected components, whereas blue bins denote the presence of undesired additional components.
    }
    \vspace{-1mm}
    \label{fig:spec_musdb_diff}
\end{figure*}

\subsection{Evaluation Metrics}
We employ objective metrics in line with the baseline system, including frequency-weighted segmental SNR (fwSNR) \cite{fwssnr}, multi-resolution spectrogram loss (MRS) \cite{mrs}, $l_1$-spectrogram distance, and Fréchet Audio Distance \cite{fad}.
Moreover, we evaluate our models using the standard signal-to-distortion ratio (SDR) metric with \texttt{mus\_eval}\footnote{\url{https://github.com/sigsep/sigsep-mus-eval}} \cite{museval}, which is widely adopted in music source separation tasks \cite{demucsv2, spleeter, openunmix}.

In addition to objective metrics, we assess our models through a subjective evaluation by conducting a Mean Opinion Score (MOS) test on Amazon Mechanical Turk.
Listeners are asked to rate the acoustic quality of the samples on a scale from 1 to 5 based on unpleasant distortions, such as noise and reverberation.
Each sample presented falls into one of the following:
1) clean music recordings or
2) clean music corrupted with random noise, reverberation, and four-frequency band equalization, as described in Section \ref{subsec:dataset}.
To evaluate the system's performance at different noise levels, we organize the listening test by using corrupted samples with SNRs of 5, 10, and 15 dB.
For reference ratings, we provided listeners with uncorrupted clean music recordings as an example of a score of 5, and identical recordings corrupted with 0dB Gaussian noise as an example of a score of 1.
In total, we collect 1,950 responses from 65 participants for the \textit{solo} dataset and 1,911 responses from 91 participants for the MUSDB18 dataset.

\section{Results}\label{sec:results}

\subsection{Objective Evaluation}\label{subsec:objeval}
Table \ref{table:obj solo musdb} presents the objective measures of \textit{Mel2Mel + Diffwave} and our proposed models on the \textit{solo} dataset, the piano-only \textit{solo} dataset, and the MUSDB18 dataset.
For the \textit{solo} dataset, our proposed models outperform the previous state-of-the-art model by a considerable margin.
The TFC-CPq model achieves significant gains on all metrics except fwSNR. 
There is a gain in SDR by 6.82dB and 6.44dB compared to the corrupted low-quality musical recordings and their enhanced outputs of \textit{Mel2Mel + Diffwave}, respectively for the TFC-CPq model.
The TFC-CPv model achieves the largest gain on fwSNR by 2.66 compared to \textit{Mel2Mel + Diffwave}.
Additionally, our models perform well on the FAD metric, which is closely aligned with perceptual quality in music enhancement tasks, according to \cite{refme}.
Among the TF-Conformer modules, TFC-PC has the worst performance on all metrics except $l_1$-spectrogram distance, where the outcome is the poorest for the TFC-C model.

We also train and evaluate our models on the piano-only \textit{solo} dataset, since there are officially available pre-trained parameters of \textit{Mel2Mel + Diffwave}. 
In this scenario, TFC-PC achieves the best results in fwSNR, $l_1$-spectrogram distance, and MRS.
From the perspective of SDR, TFC-CPq performs the best, demonstrating performance gains of 7.33dB and 5.76dB compared to low-quality samples and \textit{Mel2Mel + Diffwave}-enhanced samples, respectively.

When it comes to enhancing the mixtures of MUSDB18, which is a multi-track dataset, the Diffwave vocoder is unable to learn how to transform mel-spectrograms to waveforms, generating only Gaussian noise.
Compared with noisy inputs, we can observe for TFC-CPq an 8.05dB gain on SDR.
As in the case of the \textit{solo} dataset, TFC-CPq achieves the most significant gain on all metrics except fwSNR.
The worst performances among the TF-Conformer modules are TFC-CPv on MRS, $l_1$ spectrogram distance, and SDR, and TFC-PC on fwSNR and FAD.


Additionally, we reported the evaluation results for \textit{solo} dataset samples enhanced by models trained on the MUSDB18 dataset in the last row of Table \ref{table:obj solo musdb}. 
One can observe that the enhanced results of the single-stem \textit{solo} dataset are comparable to those of \textit{Mel2Mel + Diffwave}, even though the proposed models were trained on the multi-track mixtures of the MUSDB18 dataset, which has a different data distribution.
This implies that our proposed models can enhance more general musical recordings. 

\subsection{Analysis of Spectrograms for Enhanced Samples}
Fig. \ref{fig:spec_solo} and Fig. \ref{fig:spec_musdb_diff} display the spectrograms of low-quality, enhanced, and ground truth musical recordings, containing the outputs of TFC-CPq module which demonstrate the best performance as shown in Section \ref{subsec:objeval}.
In Fig. \ref{fig:spec_solo}, we present the outcomes for various instruments in the Medley-solos-DB dataset.
It is evident that both \textit{Mel2Mel + Diffwave} and our proposed model effectively enhance low-quality music.
In the flute example, we can observe that our model effectively denoises the background clatter noise, which impacts a wide range of frequency bands in narrow time bins, particularly in the early frames. 
The piano example demonstrates that our proposed model more elaborately recovers harmonics in higher frequencies compared to the previous state-of-the-art model. 
In the violin and clarinet examples, both models effectively eliminate noise across most frequency bands; however, our proposed model demonstrates more precise restorations of spectral features. 
This could be attributed to the benefits of using a mask-based model over a generative model. 

Figure \ref{fig:spec_musdb_diff} displays the results for the MUSDB18 dataset. 
As the Diffwave vocoder \cite{diffwave} failed to converge during training for the MUSDB18 dataset, we only report the results for our proposed model. 
The performance is comparable to that observed in the Medley-solos-DB dataset.
Example 1 demonstrates effective restoration of the low-frequency band, where the bass sound is primarily located.
It is noteworthy that examples 2 and 3 exhibits proficient enhancement performance for different types of corruption within a similar frequency band.
The low-quality audio in example 2 is corrupted by equalization, simulating the effect of a low-quality microphone. 
Our proposed model successfully recovers the corrupted spectral components. 
Example 3 presents audio corruption in a similar frequency band, in this case, with an unexpectedly higher gain.
Our model effectively suppresses this corruption as well, as evidenced by the difference in spectrograms between the low-quality and enhanced audio compared to the clean ground truth audio.
On the other hand, example 4 presents a distinct scenario in which the corrupted frequency bands lie within a lower range compared to those in examples 2 and 3, where spectral features are also successfully recovered. 
Furthermore, as depicted in example 5, our model is capable of enhancing music corrupted across both low- and high-frequency bands.

\begin{table}[t]
    \begin{subtable}[t]{0.98\columnwidth}
        \centering
        \begin{tabular}{l ccc}
        \toprule
        Model    & SNR 5  & SNR 10 & SNR15 \\
        \midrule
        HQ  & \multicolumn{3}{c}{$4.05\pm0.12$ (no noise)} \\
        \midrule
        LQ   & $2.61\pm0.16$ & $3.06\pm0.17$ & $3.30\pm0.17$\\
        \textit{M+D} & $3.62\pm0.16$ & $3.56\pm0.16$ & $3.75\pm0.15$ \\
        TFC-CPq & \textbf{3.76 $\pm$ 0.16}  & \textbf{3.73 $\pm$ 0.16} & \textbf{3.80 $\pm$ 0.15} \\
        \bottomrule
        \end{tabular}
        \caption{MOS results of \textit{solo} dataset.}
    \end{subtable}
    
    \begin{subtable}[t]{0.98\columnwidth}
        \centering
        \begin{tabular}{l ccc}
        \toprule
        Model    & SNR 5  & SNR 10 & SNR15 \\
        \midrule
        HQ  & \multicolumn{3}{c}{$4.02\pm0.09$ (no noise)} \\
        \midrule
        LQ   & $3.22\pm0.12$ & $3.50\pm0.14$ & $3.47\pm0.14$\\
        TFC-CPq & \textbf{3.62 $\pm$ 0.13} & \textbf{3.64 $\pm$ 0.11} & \textbf{3.70 $\pm$ 0.12} \\
        \bottomrule
        \end{tabular}
        \caption{MOS results of MUSDB18 dataset.}
    \end{subtable}
    \caption{Mean Opinion Scores (MOS) of \textit{solo} dataset and MUSDB18 dataset. The ratings for enhanced samples for the corresponding signal-to-noise ratio (SNR) are listed in each column, along with 95\% confidence intervals.}
    \label{tab:mos}
\end{table}

\subsection{Subjective Evaluation}
The Table \ref{tab:mos} displays the Mean Opinion Scores (MOS) of the corrupted samples of \textit{solo} dataset and MUSDB18 dataset enhanced by the TFC-CPq model, which achieved the highest scores in FAD.
For the \textit{solo} dataset, the MOS results indicate that our TFC-CPq model may have better perceptual quality than the previous \textit{Mel2Mel + Diffwave} model for all SNRs. 
Notably, for samples with 5dB SNR, TFC-CPq achieves significant MOS improvements of 1.15 for the \textit{solo} dataset and 0.40 for the MUSDB18 dataset compared to low-quality ones, respectively.
When compared to the \textit{Mel2Mel + Diffwave} model, TFC-CPq has an MOS gain of 0.06, 0.17, and 0.05 for SNRs of 5dB, 10dB and 15dB, suggesting that TFC-CPq may have a tendency to produce more perceptually pleasing samples. 

Even for samples with 15dB SNR, where the presence of additive noise is relatively low, TFC-CPq scored 0.5 and 0.23 higher MOS compared to low-quality samples for both datasets, implying that our models can enhance the musical recordings with improved reverberation and equalization from a perceptual standpoint.
These results suggest that our proposed TFC-CPq model could potentially be effective in enhancing the subjective quality of music recordings across various SNRs.
For the MUSDB18 dataset, the MOS results also show that our TFC-CPq model outperforms the low-quality baseline. Specifically, for samples with 5dB SNR, TFC-CPq achieves an MOS improvement of 0.40 compared to low-quality samples.

Despite these encouraging results, our models have some limitations.
They often fail to denoise impulses, such as the sound of objects colliding.
Additionally, since they recognize the babbles as vocal, our models tend to sharpen rather than erase them. 
We will address these issues in future work.

In summary, the subjective evaluation results support the findings of the objective evaluation, demonstrating that our proposed TFC-CPq model achieves superior perceptual quality compared to the \textit{Mel2Mel + Diffwave} model and the low-quality baseline across various SNRs and datasets. The consistent performance of our model in both \textit{solo} and MUSDB18 datasets suggest that our proposed methods can generalize well to different types of musical recordings.

\section{Conclusions}\label{sec:conclusions}
In this study, we proposed the use of TF-Conformers to perform music enhancement tasks, which demonstrated significant improvements in performance compared to existing methods. 
Furthermore, we expanded our investigation to multi-track musical recordings, an area previously unexplored in the literature.
In future works, our research will explore additional applications, such as music source separation and speech enhancement.
Additionally, we plan to apply our method to larger real-world datasets, such as YouTube recordings, by utilizing unsupervised techniques to further advance the field of music enhancement.

\section{Acknowledgements}
\label{section:ack}
This work was partly supported by Institute of Information \& communications Technology Planning \& Evaluation (IITP) grant funded by the Korea government(MSIT)[NO. 2022-0-00641, XVoice: Multi-Modal Voice Meta Learning, 50\%], [NO.2021-0-01343, Artificial Intelligence Graduate School Program (Seoul National University), 10\%], and Culture, Sports and Tourism R\&D Program through the Korea Creative Content Agency grant funded by the Ministry of Culture, Sports and Tourism in 2022 [No.R2022020066, 40\%].

\newpage



\bibliographystyle{ACM-Reference-Format}
\balance
\bibliography{ref}


\end{document}